\documentclass[journal=ancac3,manuscript=article]{achemso}
\setkeys{acs}{articletitle = true}
\usepackage[version=3]{mhchem} % Formula subscripts using \ce{}
\usepackage{amsmath}
\usepackage{color}
\usepackage[normalem]{ulem}
\setkeys{acs}{
  abbreviations = true,
  maxauthors    = 20,
}

\author{Emanuele Galiffi}
\affiliation{Imperial College London, Department of Physics, The Blackett Laboratory, London SW7 2AZ, UK}
\email{emanuele.galiffi12@imperial.ac.uk}
\author{John B. Pendry}
\affiliation{Imperial College London, Department of Physics, The Blackett Laboratory, London SW7 2AZ, UK}
\author{Paloma A. Huidobro}
\affiliation{Imperial College London, Department of Physics, The Blackett Laboratory, London SW7 2AZ, UK}

\title[Broadband Tunable THz Absorption with Singular Graphene Metasurfaces]
  {Broadband Tunable THz Absorption with Singular Graphene Metasurfaces}


\begin{document}


% \begin{tocentry}

% Graphene recently earned itself a spotlight in plasmonics, thanks to its high tunability and low losses, which enable the design of efficient metasurfaces by dynamical doping. In this work we present a graphene metagrating which acts as a strong broadband THz absorber, thanks to the presence of a point of very low carrier density. Its working frequency, which lies in the technologically relevant low THz region, may be tuned by adjusting the grating periodicity, and the doping can be potentially modulated at high frequencies. 

% \end{tocentry}

\begin{abstract}

By exploiting singular spatial modulations of the graphene conductivity, we design a broadband, tunable THz absorber whose efficiency approaches the theoretical upper bound for a wide absorption band with a fractional bandwidth of 185\%. Strong field enhancement is exhibited by the modes of this extended structure, which is able to excite a wealth of high order surface plasmons, enabling deeply subwavelength focussing of incident THz radiation. Previous studies have shown that the conductivity can be modulated at GHz frequencies, which might lead to the development of efficient high speed broadband switching by an atomically thin layer.

\end{abstract}
% * <emaciao.92@gmail.com> 2017-12-14T17:18:06.970Z:
% 
% Replaced "remarkably" with the actual bandwidth of the device. Replaced "opening new possibilities" by "which might lead to the development of"
% 
% ^.


When suitably doped, graphene supports THz plasmons\cite{Basov2016,Low2017}, and these can be coupled to incident radiation by periodic modification of the sample either by doping\cite{Fang2013,Poumirol2017} or by physical restructuring.\cite{Yan2012} Very strong absorption results, approaching the theoretical limit of $50\%$, but over an extremely narrow range of frequencies, were already demonstrated.\cite{Huidobro2017a} In a previous paper\cite{Pendry2017} we have shown that singular metasurfaces are qualitatively different from non singular ones. Here we exploit the broadband nature of their spectrum to create an absorber of THz radiation with equally strong absorption but now over a broad band of frequencies and, curiously, in a layer of material a fraction of a nanometre thick. The absorption has the potential to be switched at a high rate, potentially as high as GHz.\cite{li2014ultrafast}
% * <emaciao.92@gmail.com> 2017-12-14T17:23:42.052Z:
% 
% replaced "remarkably" by "curiously"
% 
% ^.
Graphene is a promising platform for nano-optics due to its high sensitivity to external electromagnetic (EM) fields\cite{Bonaccorso2010,Low2014}. Despite being atomically thin, this two-dimensional (2D) material is able to sustain oscillations of its free electrons driven by EM radiation, thanks to the excitation of surface plasmon polaritons.\cite{Vafek2006,Hanson2008,Jablan2009,Dubinov2011,
Koppens2011,Nikitin2011,Fei2011} Its high, doping-independent, carrier mobility, \cite{Ni2016,chen2008intrinsic} and the possibility of electrical tuning of its Fermi level,\cite{Vakil2011,Novoselov2004} make graphene an ideal material for plasmonic applications.\cite{GarciadeAbajo2014,Grigorenko2012} 

Complete light absorption has formerly been realized in a variety of implementations, most often, but not only, exploiting interference effects following the Salisbury screen concept. Multi-layer structures~\cite{Huidobro2017a,nefedov2013perfect}, but also periodic surface patterning~\cite{zhang2014coherent} and other metamaterial configurations~\cite{thongrattanasiri2012complete,fang2012tunable} have shown good results over a narrow absorption band. More recently, the tunability of graphene inspired applications of this material in the design of absorptive metasurfaces.\cite{fang2013active,Huidobro2017a} A general theoretical framework for the design of coherent broadband absorbers was recently investigated using scattering matrices.\cite{kim2016general}. In the near-infrared, perfect broadband absorbers have been recently realized in epsilon-near-zero~\cite{liberal2017near} indium-tin-oxide stacks, achieving $>50\%$ absorption over a fractional bandwidth of $\approx0.5$.\cite{yoon2015broadband}. In an $\epsilon < 0$ material, broadband response is characteristic of singular structures. Localized plasmonic structures with sharp edges or touching points have been studied in the past.\cite{Luo2010,Pendry2013,aubry2010broadband} In these systems, radiation is captured in their smooth portions, and travels towards the singularities becoming increasingly squeezed. This leads to a continuous rather than discrete far field spectrum. In this work we show such a broadband response for an extended metasurface, rather than for a localized structure.

Applications in plasmonics rely on the capacity of concentrating and delivering electromagnetic energy at the nanoscale.\cite{Maier2007Plasmonics} The large spatial confinement provided by graphene plasmons is highly desirable for this purpose, but at the same time it limits our ability to access and tailor these modes. This has been sorted in the past by using local probes,\cite{Fei2012,Chen2012,Alonso-Gonzalez2014,Alonso-Gonzalez2016} or by making use of subwavelength periodic arrays of patterned graphene structures and gratings\cite{Zhan2012,Nikitin2012b,Nikitin2012,
Thongrattanasiri2012a,Slipchenko2013,Peres2013,
Stauber2014a,Miao2015,Zhao2015}. 
Periodic arrays and gratings have also been proposed as metasurfaces, which enable the manipulation of radiation with a single atomic layer in the technologically relevant THz regime.\cite{Ju2011,Tassin2013,Fan2015,Li2015,Huidobro2016a}

Here we go one step further and consider a singular graphene metasurface: a graphene sheet is periodically doped along one spatial dimension to form a subwavelength grating with vanishing Fermi level at the minimum grating points. As we have recently shown,\cite{Pendry2017} these graphene gratings are an example of singular plasmonic metasurfaces whose spectral properties are inherited from a higher dimensional structure, such that they compress one or more dimensions of the original space into the singularities. This results in metasurfaces whose spectral response is determined not just by the wave vector of incident radiation as for a conventional grating, but also by an extra, hidden, wave vector, inherited from the mother structure. As a consequence, the modes form a continuum and this way we design a graphene metasurface with broadband THz absorption. The underlying physics is elegantly illuminated by our transformation optics approach, which enables us to relate the spectrum of our singular metasurface to that of a simple translationally invariant slab with the desired spectral features.

This paper is organized as follows. We start by using transformation optics \cite{Ward1996,Pendry2006,Leonhardt2006} to design a singular plasmonic metasurface, and we discuss how its broadband spectrum arises. We then show how periodically doped graphene can serve as a platform for such singular metasurfaces. We present analytical results for the transmittance and absorbance through these metasurfaces, and detail the effect of approaching the singular limit. We show how a single graphene layer doped in the prescribed way presents strong broadband absorption in contrast to the series of discrete absorption peaks of conventional gratings. The effect of losses in the performance of this device is then discussed, and we conclude by demonstrating the field enhancement achieved by our metasurface.

\section{Results and discussion}

\begin{figure}[ht!]
\begin{center}
\includegraphics[angle=0, width=0.98\textwidth]{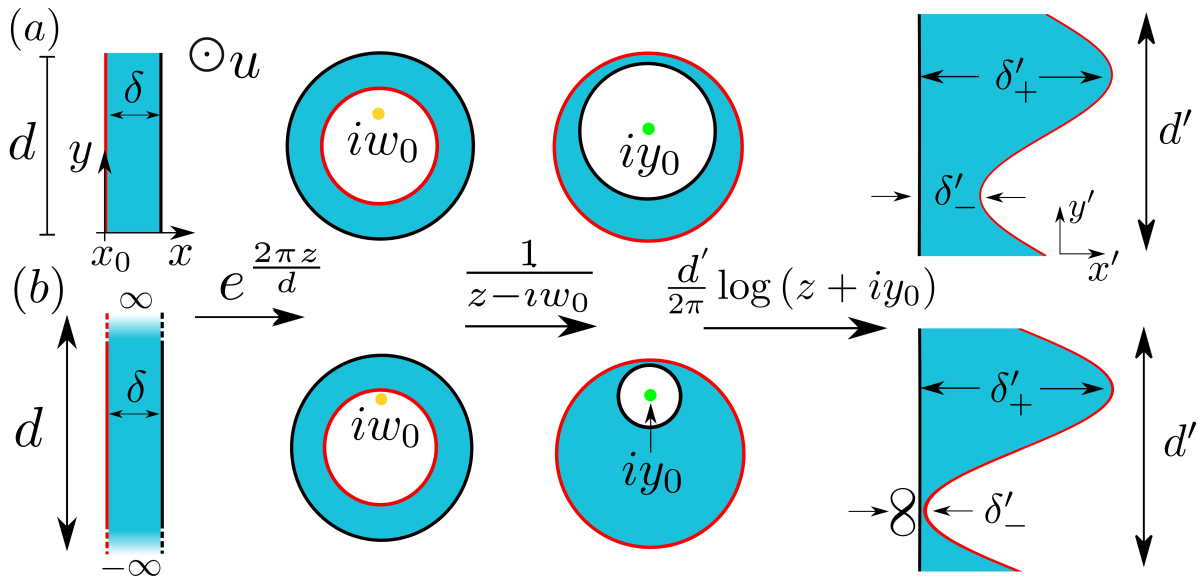}
\caption[]{Generating a conventional (a) and singular (b) grating with conformal transformations. (a) A flat plasmonic slab is mapped onto a plasmonic film of periodically modulated thickness by a succession of transformations. Parameters $w_0$, $y_0$, $d$ and $\delta$ control the shape of the grating. (b) As $w_0$ increases towards 1, the resulting grating approaches the singular limit, where the thickness of the grating vanishes at the touching points.}\label{FigTransformation}
\end{center}
\end{figure}

A series of conformal transformations can be used to design a plasmonic grating \cite{Kraft2015} as depicted in Figure \ref{FigTransformation}a. We start from a simple flat vertical slab of plasmonic material in frame $z=x+iy$, which we refer hereafter as slab frame (see Figure~\ref{FigTransformation}a). The slab has permittivity $\epsilon(\omega)$ and is surrounded by dielectrics.  %$\epsilon_1$ and $\epsilon_2$. 
It is placed at $x=x_0$ (we take $x_0=0$ without loss of generality), has thickness $\delta$, and is translationally invariant along the $y$ direction, as well as along $u$, the out of plane dimension. Its modes are characterized by two wavevectors, $k_y$ and $k_u$. Each finite, invariant slab of length $d$ of this infinite structure is mapped to a concentric annulus through $e^{2\pi z/d}$, and is then off-centered by inverting it about the point $i w_0$, which lies inside the annulus. Finally, a $\log$ transformation results in an infinite slab in the grating frame, $z'=x'+iy'$. The periodicity $d'$ of the grating is fixed \textit{via} the final prefactor $\frac{d'}{2\pi}$. In this frame, the slab thickness, $\delta'$, is periodically modulated along $y'$ with period $d'$. The cascaded transformation is given by,

\begin{equation}
	z'=\frac{d'}{2\pi }\log\left( \frac{1}{e^{2\pi z/d} - i w_0} + i y_0 \right), 
	\label{eq:transformation}
\end{equation} 
where $y_0 = w_0/\left[e^{4\pi(x_0+\delta)/d}-w_0^2\right]$ is the distance of the center of the inner circle of the inverted, non-concentric annulus from the origin.

Singular plasmonic gratings can be generated as follows. As the inversion point $i w_0$ approaches the inner cylinder in Figure \ref{FigTransformation}b, the maximum modulation amplitude $\delta_+'$ increases to a larger value $\bar{\delta}_+'$, which diverges in the limit $w_0\rightarrow \exp(x_0)=1$. However, we require the thickness at the peak point $\delta_+'$ and period $d'$ of the grating to be fixed. As explained in the Supporting Information (SI), these constraints are satisfied by simultaneously increasing the length $d$ of the invariant slab by a factor $\frac{\bar{\delta}_+'}{\delta_+'}$. Hence, in order to approach the singular limit, we need to let $d \rightarrow \infty$, such that the points of vanishing thickness in the grating frame are mapped into $\infty$ in the slab frame, as shown in Figure \ref{FigTransformation}b. This generates a set of gratings for different values of $w_0$ that, in the grating frame, only differ in their minimum thickness, $\delta'_-$, which vanishes in the singular limit. The effect thus produced is surprising: since the finite grating inherits its spectral response from the infinite, translationally invariant mother structure, our singular grating acquires the same broadband plasmonic spectrum. In a different picture, although the singular grating appears as a 2D system, it is indeed characterized by three wave vectors: the wave vector in the grating coordinate system, $k_{y'}$, the out of plane one, $k_u$, and a hidden wave vector associated with the extra dimension inherited from the original infinite slab, $k_y$.

\begin{figure}[t!]
\begin{center}
\includegraphics [width=0.5\textwidth]{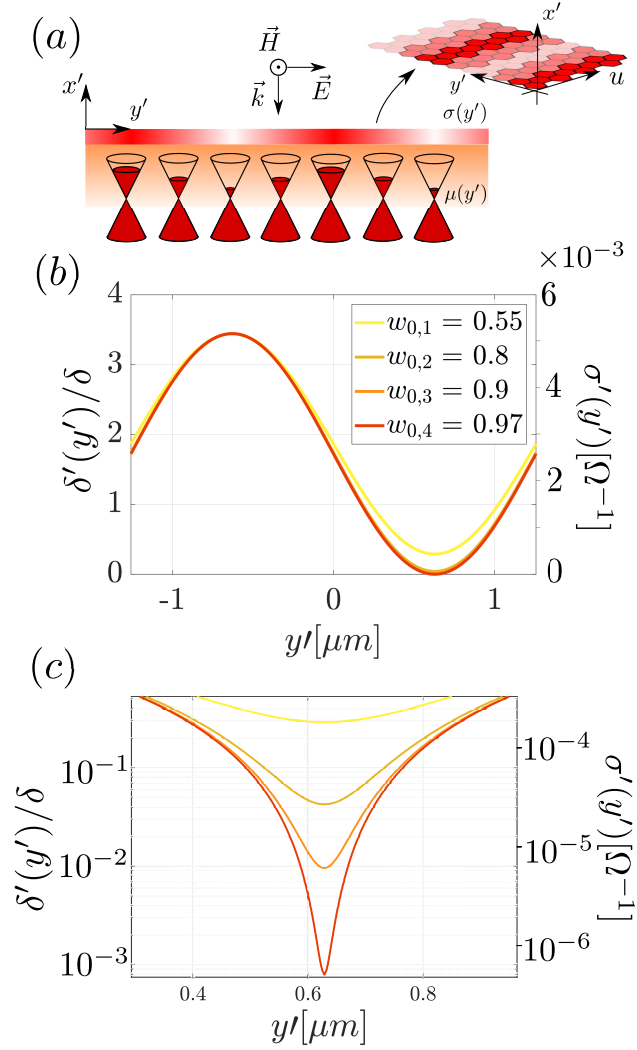}
\caption[]{Graphene metasurface with periodically modulated conductivity. (a) The thin grating designed with conformal transformations shown in Figure \ref{FigTransformation} is used to model a graphene sheet with periodically modulated doping level.  
(b) Grating shape over one period (2.5 $\mu$m) as the singular limit is approached by increasing $w_0$ and $d$ simultaneously (the corresponding values of $d$, in units of $2\pi$, are: 1, 16.32, 34.15 and 113.01). The left axis represents the thickness ratio between grating and slab frame, $\delta'(y')/\delta$, while the right axis gives the conductivity profile $\sigma'(y')$ for a representative frequency, $6.0$ THz.
(c) Zoom of panel (b) close to the singular point. The carrier's mobility is taken as $m=10^4$ cm$^2/$(V$\cdot$s).}\label{GrapheneGrating}
\end{center}
\end{figure}

As we have previously demonstrated~\cite{Huidobro2016}, the plasmonic grating shown in the right hand side of Figure~\ref{FigTransformation} represents a graphene sheet with periodically modulated conductivity (see sketch in Figure \ref{GrapheneGrating}a if the periodicity is much larger than the thickness, $d'\gg \delta'$. In the slab frame, the original system is a graphene sheet with homogeneous conductivity, $\sigma(\omega)$,  which we take from the random phase approximation as detailed in the Methods section and depends on chemical potential $\mu$, carrier's mobility $m$ and temperature $T$ (we use $T=300$ K). Working with 2D conformal transformations conserves the in-plane permittivity $\epsilon$ and permeability $\mu$ of materials between the different frames. Hence, the permittivity of the homogeneously doped graphene, $\epsilon(\omega)=1+i\sigma(\omega)/(\omega\epsilon_0\delta)$, is conserved when we transform to the periodically doped one.\cite{Leonhardt2006} This implies that the conductivity of the graphene metasurface is modulated along the $y'$ coordinate following,
\begin{equation}
	\sigma'(y',\omega) = \frac{\delta'(y')}{\delta}\sigma(\omega). 
	\label{eq:modulatedsigma}
\end{equation}  
In order to investigate the singular metasurfaces, we start with a case away from the singular limit, $w_0=0.55$, for which we set $d=2\pi$. As we allow $w_0$ to increase towards 1, we renormalize the entire grating thickness to match the peak modulation by extending the slab length $d$ as outlined above. This design procedure ensures that all the gratings considered keep the same peak modulation, $\delta'_+$, while the minimum modulation, $\delta'_-$ is gradually reduced. Figure \ref{GrapheneGrating}b shows the modulation profile, $\delta'(y')/\delta$, for a chosen grating period ($d'=2\pi\cdot400$ nm $\approx2.5$ $\mu$m) and different values of the modulation parameter, $w_0$. The thickness reduction at the singular point is shown in detail in panel c. 

From Eq.~\ref{eq:modulatedsigma}, we see that rescaling $\delta'$ is equivalent to rescaling the spatially constant conductivity in the slab frame. At the same time, the modulated conductivity $\sigma'$ depends also on the frequency through the conductivity in the slab frame, $\sigma(\omega)$. The conductivity modulation at a representative frequency of 6 THz is presented in panel b (right hand axis) for the different values of $w_0$ considered in Figure \ref{GrapheneGrating}. %The chemical potential and carrier's mobility are fixed in the slab frame as $\mu=0.5$ eV (for $w_0=0.55$) and $m=10^4$ cm$^2/$(V$\cdot$s) (for all cases). 
The peak conductivity corresponds to a maximum value of $\mu=1.6$ eV. On the other hand, since there is no linear relationship between $\mu$ and $\sigma$ outside of the Drude limit, \textit{i.e.} for $\mu\ll\hbar\omega$, the chemical potential values needed at the minimum point are not extremely low. In the SI we provide a detailed discussion of this regime and discuss the feasibility of these metasurfaces.

Transformation optics, which exploits the invariance of Maxwell's equations under coordinate transformations \cite{Ward1996,Pendry2006,Leonhardt2006}, can be used to analytically calculate the response of these graphene metasurfaces under a p-polarized EM wave at normal incidence (magnetic field $\mathbf{H}=H\mathbf{u}$, $k_u = 0$).\cite{Huidobro2016} Making use of the fact that 2D conformal maps preserve the scalar potential we transform the source plane wave from the grating frame to the slab frame and apply the boundary conditions for the EM field in that frame, where the conductivity of graphene is homogeneous. In a second step, we include the radiative reaction from the grating. Due to the fact that graphene confines plasmons to an extremely subwavelength regime, including radiative losses perturbatively yields very accurate analytical results. Our analytical approach thus allows us to accurately model gratings that are very close to the singular limit in a computationally efficient way, as opposed to carrying out a full numerical treatment.

\begin{figure}[t!]
\begin{center}
\includegraphics [angle=0, width=0.95\textwidth]{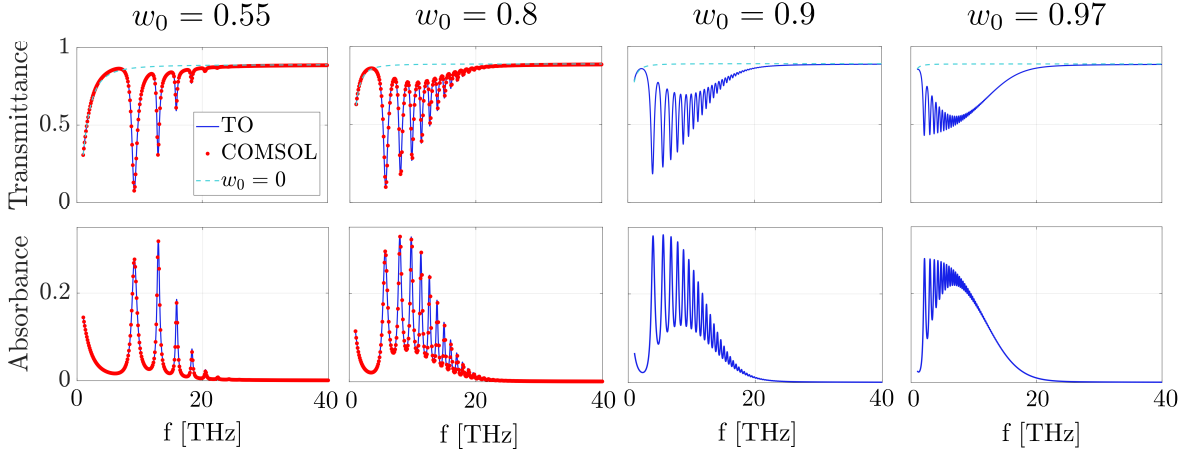}
\caption[]{Merging the normal incidence spectrum of graphene gratings into a continuum. Transmittance (top row) and absorbance (bottom row) for different values of $w_0$ are shown as the grating approaches the singular limit. The modulation profile of each grating correspond to the ones presented in Figure \ref{GrapheneGrating}. Analytical results are depicted with blue lines and numerical results with red dots. The graphene metasurfaces are placed on a substrate of refractive index $n=2$, that could model SiO$_2$ in the THz regime. }\label{FigTransAbs}
\end{center}
\end{figure}

Figure \ref{FigTransAbs} shows the analytical transmittance (top row) and absorbance (bottom row) spectra for a set of graphene metasurfaces with the modulation profiles discussed in Figure \ref{GrapheneGrating}. A semi-infinite substrate with permittivity $\Re{[\epsilon_2]}=4$ is assumed, which corresponds to the average value for SiO$_2$ in this frequency range. We also present numerical results from full electrodynamic simulations (see Methods) for the two least singular cases, which show the accuracy of our analytical method (see red dots in Figure \ref{FigTransAbs}). Let us first focus on a case far from the singular limit, $w_0=0.55$. The spectrum shows a discrete series of well separated resonances (transmission dips and absorption peaks), each of them corresponding to the excitation of a plasmonic mode. The response of this metasurface is that of a conventional subwavelength grating: it has a discrete spectrum defined by two wavevectors, the parallel momentum along the grating, $k_{y'}$, and the out of plane momentum ($k_u=0$ in our study), which are fixed by the incident radiation. In other words, conventional gratings are coloured. The periodicity of the grating defines the set of modes, with increasing multipolar orders lying at higher frequencies. As the mode order increases, the radiative coupling of the resonance modes decreases and so does the contrast of transmittance peaks. On the other hand, the highest radiative coupling of the lowest order modes results in an increased absorption, such that the highest absorption peak occurs mid spectrum and approaches the theoretical upper bound $A_{\text{max}}=1/(1+\sqrt{\epsilon_2})\approx 0.33$. In previous works we have shown how these resonances can be used to design a tunable plasmonic metasurface \cite{Huidobro2016a} or a narrowband perfect absorber \cite{Huidobro2017}, focusing on the first resonance mode (dipolar).

We now focus on the behaviour of the metasurface as the singular limit is approached. The physics at play is best illuminated by analysing the spectrally equivalent mother structure in the slab frame. As already explained, a singular grating is generated from an invariant slab of length $d\to \infty$. This structure has a continuous spectrum. Hence, the width of the Brillouin zone in the slab frame is reduced, meaning that the resonances surface plasmon resonances merge into a continuum of modes. In other words, a singular metasurface is spectrally black.

At the same time, as explained above and in the SI, extending the slab periodicity $d$ has the effect of scaling down the modulated thickness $\delta(y')$ of the grating by the same factor. This is equivalent to rescaling the constant conductivity in the slab frame according to Eq.~\ref{eq:modulatedsigma}. From the dispersion relation in the quasistatic approximation\cite{Huidobro2016a}:
\begin{equation}
    \epsilon_1 + \epsilon_2 + \frac{4\pi i \sigma(\omega)}{\omega} |k| = 0 \label{DispRel}
\end{equation}
where $\omega$ is the angular frequency and $|k|$ the in-plane wavevector,  it is straightforward to show that reducing $\sigma$ by a constant has the effect of lowering the resonances. In fact, assuming a Drude conductivity, the resonance frequencies are proportional to the square root of the conductivity, which explains the reduction in the resonance frequencies to the low THz region.

These effects are already visible for the case $w_0=0.8$. In addition, the radiative broadening of the peaks is reduced as the singularity is approached: as the weight of the plasmon is drawn into the singularity, its coupling to the outside world is reduced. Hence, more peaks are visible in the transmittance (top row) and absorbance (bottom row) spectra compared to $w_0=0.55$. Increasing $w_0$ to 0.9 further lowers the resonance frequencies and radiative broadening, and the peaks get closer and closer together. Finally, for $w_0=0.97$, the higher order peaks are merging into a continuum as their spacing becomes less than the resistive broadening. This is the singular limit, where the hidden dimension in the singular plasmonic metasurface dictates its response to external radiation:\cite{Pendry2017} rather than a discrete set of peaks fixed by the incident wave vector $k_y'$, the spectrum is broadened into a continuum. Merging the spectrum results in a broad band where absorption is close to its maximum possible value.

On the other hand, it is well known that nonlocality has relevant effects in the spectrum of plasmonic nanostructures with sharp features over very short distances.\cite{Savage2012,Ciraci2012} Nonlocality becomes relevant at distances comparable to the Fermi wavelength of electrons, where the screening charge is smeared over the Thomas-Fermi screening length, rather than being localized at the surface of the plasmonic structure. For instance, the gap between two nearly touching plasmonic structures will be effectively increased by such charge smearing thus limiting the field enhancement from increasing up to infinity as classical theories would predict.\cite{Luo2013} Similarly, nonlocality will limit the ability of our singular graphene metasurfaces to concentrate radiation at singular points. Nonlocal screening of electron charges close to the minimum conductivity region will smear out the singularity, which will prevent the field enhancement from growing to infinity and the graphene plasmons from travelling at very low group velocities close to the Fermi velocity.\cite{Lundeberg2017} This will effectively tend to produce a discrete rather than continuous spectrum, removing the possibility of a perfect singularity. At the same time, higher resistive losses will tend to smear out the spectrum into a continuum. As we show below, higher losses result in a complete smearing of the spectrum, thus merging the discrete peaks into a continuum. Hence losses and non-locality will predictably tend to compensate one another.

\begin{figure}[t!]
\begin{center}
\includegraphics[angle=0, width=0.5\textwidth]{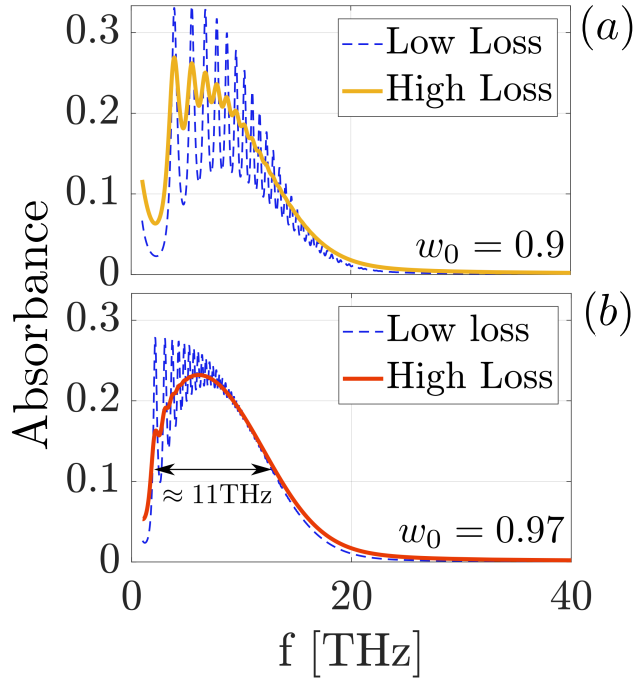}
\caption[]{Effect of losses in the spectrum of singular graphene metasurfaces, demonstrating how the peaks are completely merged into a continuum once the resistive broadening reaches the spacing between the modes. Absorbance is shown for the two most singular metasurfaces considered in Figure \ref{FigTransAbs} for different values of losses. Carrier mobilities used are: $m=10^4$ (dashed line) and $3\times10^3$ cm$^2/$(V$\cdot$s) (continuous line), corresponding to electron scattering frequencies of $1.8$ and $6.0$ THz respectively. The full band width at half maximum is $\approx 11.1$ THz.}\label{FigLosses}
\end{center}
\end{figure}

The effect of losses in the spectrum of the two most singular gratings is shown in Figure \ref{FigLosses}, where the carrier's mobility is reduced from $10^4$ to $3\times10^3$ cm$^2/$(V$\cdot$s), which are all experimentally achieved values.\cite{chen2008intrinsic,wu2009half} While all the main features characterising these metasurfaces remain, the consequence of lower carrier mobilities is to increase the non-radiative broadening of the peaks and hence reduce the transmittance and absorption peak contrasts. The effect is stronger for the higher order peaks, and results in a quicker merging into the continuum for lower values of the modulation strength $w_0$, that is, further away from the singular limit. This suggests that the singular graphene metasurface can act as a compact and broadband THz absorber, even for lower quality graphene samples, which is of great relevance for THz applications.\cite{Tonouchi2007} This effect is in contrast with usual devices based on the Salisbury screen concept, which are narrowband.

The bandwidth of our graphene absorber may be calculated as the full width at half maximum of the red curve in Figure~\ref{FigLosses}. The maximum absorbance achieved by our most lossy singular metasurface at $f_{max}\approx 6.0$ THz, is $0.232$, corresponding to $\approx 70\%$ of the theoretical upper bound on an SiO$_2$ substrate. The half maximum frequencies are $f\approx1.8$ THz and $f\approx12.9$ THz, resulting in a fractional bandwidth $\frac{\Delta f}{f_{max}} \approx 1.85$. In addition, the working frequency of the device may be tuned by adjusting the periodicity of the grating. The square root dependence of the resonance frequencies on the width of the Brillouin zone can be easily seen from the dispersion relation (Eq.~\ref{DispRel}). Surprisingly, such large, broadband, tunable THz light harvesting occurs in an atom-thick layer of plasmonic material.

\begin{figure}[t!]
\begin{center}
\includegraphics [angle=0, width=0.7\textwidth]{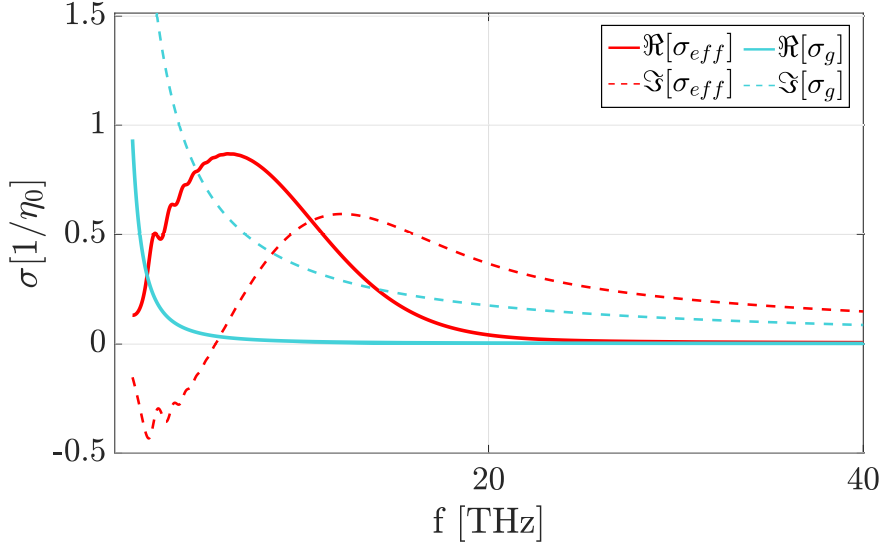}
\caption[]{Our formalism enables the calculation of a homogenized conductivity for the metasurface. The real (solid lines) and imaginary (dashed lines) parts of the effective conductivity $\sigma_{eff}$ (red lines), as well as the original, homogeneous graphene conductivity $\sigma_g$ (light blue lines) in the singular limit ($w_0=0.97$) are shown for higher losses in units of $1/\eta_0$, where $\eta_0 \approx 377$ $\Omega$ is the impedance of free space. }\label{FigSigEff}
\end{center}
\end{figure}

Since the period of the modulation is much smaller than the operating wavelength, $d'\ll\lambda$, the response of the graphene metasurface can be described through an effective conductivity.\cite{Huidobro2016a} 
This is plotted in Figure \ref{FigSigEff} in the singular limit, together with the conductivity of a homogeneously doped graphene sample. The difference between the two is striking: as shown previously\cite{Huidobro2016a}, a conventional metasurface would behave as an absorber at the individual resonance peaks. However, in the singular limit the peaks overlap, so that the minimum value of $\text{Re}[\sigma_{\text{eff}}]$ between peaks increases above the background value, as evident from the low loss plot. Hence, the resonances are fully merged into a wide, continuous absorption band.

\begin{figure}[t!]
\begin{center}
\includegraphics [angle=0, width=0.9\textwidth]{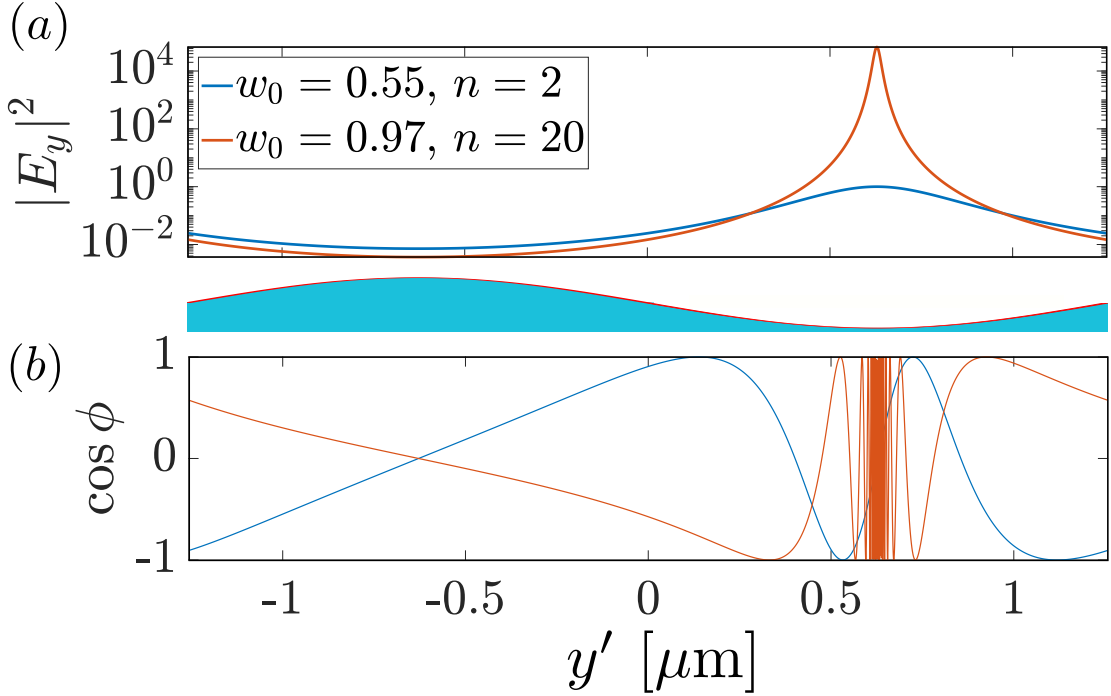}
\caption[]{ Field plots calculated analytically at the graphene position. (a) Square modulus of the in-plane component of the electric field. The blue curve corresponds to the $n=2$ mode (the second peak in the absorption spectrum), with eigenfrequency $f\approx 12.93$ THz (see Figure \ref{FigTransAbs}), of the non-singular grating ($w_0 = 0.55$). The red curve corresponds to the $n=20$ mode of the singular ($w=0.97$) case, with frequency $f\approx 9.55$ THz. Both wavefunctions are normalized to the maximum value for the non-singular grating. This clearly shows a large enhancement of the field intensity by four orders of magnitude in the singular metasurface. (b) The phase of the wavefunction in the two cases aforementioned. In the singular limit, the local density of states is greatly amplified close to the singular point.}
\label{FigFieldPlot}
\end{center}
\end{figure}

The merging of the spectrum for the singular metasurfaces is accompanied by greatly enhanced EM fields. Interestingly, these metasurfaces are able to efficiently harvest THz radiation, which then travels towards the regions where the conductivity of the graphene sheet is minimum, leading to electromagnetic fields that are increasingly squeezed towards the singular point but which never reach it due to increasingly low group velocities. In particular, our singular metasurface is able to excite a wealth of higher order modes, compared to a conventional grating. Low order modes were extensively studied in a previous work, showing a moderate field enhancement of order $30$ in amplitude.\cite{Huidobro2016a} We now show how a singular metasurface ($w_0=0.97$) performs, compared to the weakly modulated one ($w_0=0.55$). A normalized field map for the non-singular case is included in the SI for reference. The surface plasmon modes of the subwavelength grating can be easily obtained by solving the Laplace equation in the slab frame, and subsequently mapping the solution onto the grating frame via the transformation rule $\vec{E}'=[\bar{\bar{\Lambda}}^{T}]^{-1}\vec{E}$, where $\bar{\bar{\Lambda}}$ is the Jacobian matrix of the transformation. As shown in Figure \ref{FigFieldPlot}a, these modes exhibit a field enhancement more than four orders of magnitude stronger in intensity $|E_y|^2$ than a non-singular grating at the graphene position. In addition, the sharp oscillations of the wavefunction near the singular point (Figure \ref{FigFieldPlot}b) imply an enormous local density of photonic states, although the finite losses in the graphene prevent the field enhancement from rising to infinity at the singular point, and the group velocity from reaching 0. Such large EM fields can be used for spectroscopies or to enhance nonlinear phenomena by placing a nonlinear material close to the graphene. Notably, the field concentration occurs in an extended structure, so the non-linearities could be enhanced with high cross section.
% * <emaciao.92@gmail.com> 2018-01-04T12:31:44.739Z:
%
% ^.
% * <emaciao.92@gmail.com> 2017-12-14T17:25:27.396Z:
% 
% Replaced "remarkably" by "notably"
% 
% ^.

\section{Conclusions}

In this work we have demonstrated how efficient collectors of THz light over a broad range of frequencies may be realized in graphene by patterning its Fermi level with a grating profile, while keeping very low carrier densities at a single point. By means of a suitable singular conformal map, transformation optics allowed us to exploit the hidden symmetries of this metasurface in order to unveil its continuous spectral response by considering an equivalent structure, consisting of a simple invariant slab. We have shown analytically how these gratings can strongly couple incident radiation into extremely confined surface plasmon modes, achieving absorption levels higher than $70\%$ of the theoretical maximum over a frequency band whose width is $185\%$ of the central frequency. These achievements promise to grant singular graphene metasurfaces a prominent role in the development of THz technology.

\section{Methods}

\subsection{Graphene's conductivity}

We take the conductivity of graphene as given by the random phase approximation, which depends on frequency $\omega$, chemical potential $\mu$, temperature $T$ and carrier's mobility $m$. It can be expressed as a sum of intraband and interband contributions, $\sigma=\sigma_{\text{intra}}+\sigma_{\text{inter}}$, as follows \cite{wunsch2006} 
\begin{eqnarray}
	\sigma_{\text{intra}} &=&\frac{2ie^2t}{\hbar\pi\left[\Omega +i\Gamma\right]}\,\text{ln}\left[ 2\,\text{cosh}\left( \frac{1}{2t} \right)\right], \label{grapheneConductivity1} \\
	\sigma_{\text{inter}}&=&\frac{e^2}{4\hbar} \left[ \frac{1}{2} +\frac{1}{\pi} \text{arctan} \left( \frac{\Omega-2}{2t} \right) - \frac{i}{2\pi} \text{ln} \frac{(\Omega+2)^2}{(\Omega-2)^2 +(2t)^2} \right]. \label{grapheneConductivity2}
\end{eqnarray}
where $\Omega=\hbar\omega/\mu$ and $t=k_BT/\mu$ are frequency and temperature normalized to the chemical potential, respectively. The normalized damping term is $\Gamma=\hbar/(\mu\tau)$, where $\tau=m\mu/v_F^2$ is the carriers' scattering time ($v_F=9.5\times 10^{7}$cm$\cdot$s$^{-1}$ is the Fermi velocity).

\subsection{Transformation Optics}
We use transformation optics to calculate analytically the reflection and transmission coefficients of a graphene sheet with periodically modulated conductivity. We consider a p-polarized plane wave incident on the graphene grating, and transform the scalar potentials to the slab frame through the conformal transformation shown in Figure \ref{FigTransformation}. The EM response of the metasurface is obtained by imposing the appropriate boundary conditions and including the radiative reaction of the graphene conductivity grating. This procedure leads to the following reflection and transmission coefficients, 
\begin{eqnarray}
	r&=&\frac{\sqrt{\epsilon_2}-\sqrt{\epsilon_1}-\eta_0\sigma_{\text{eff}}(\omega)}{\sqrt{\epsilon_2}+\sqrt{\epsilon_1}+\eta_0\sigma_{\text{eff}}(\omega)}, \\ 
	t&=&\frac{2\sqrt{\epsilon_2}}{\sqrt{\epsilon_2}+\sqrt{\epsilon_1}+\eta_0\sigma_{\text{eff}}(\omega)},
\end{eqnarray}
where $\epsilon_1$ and $\epsilon_2$ are the permittivities of air and the substrate, respectively, $\eta_0$ is the impedance of free space and $\sigma_{\text{eff}}(\omega)=\sigma(\omega)N_{\text{TO}}$ is the effective conductivity of the metasurface. Here, $\sigma(\omega)$ is the conductivity of graphene in the slab frame and $N_{\text{TO}}$ contains a sum over Fourier components of the transformation given by Eq. \ref{eq:transformation}. Full details on this derivation and the expression for $N_{\text{TO}}$ can be found in Ref. \citenum{Huidobro2016}. Transmittance and reflectance are obtained from 
\begin{eqnarray}
	T&=&\frac{\sqrt{\epsilon_1}}{\sqrt{\epsilon_2}}|t|^2, \\
	R&=&|r|^2,
\end{eqnarray}
and absorbance is $A=1-R-T$. This procedure is valid at normal incidence where all the fields are continuous through the branch points of the transformation. Out of normal incidence the waves interact with branch points and an alternative analytical approach is needed to calculate the modes of the structure.\cite{Huidobro2017a}

\subsection{Numerical simulations}
Numerical simulations were carried out with a commercial finite element method solver (Comsol Multiphysics). In our 2D simulations, the graphene layer is modelled as a current sheet with surface current $\mathbf{J}=\sigma' \mathbf{E}_{\|}$. The modulated conductivity, $\sigma'(y',\omega)$, is obtained from Eq.~\ref{eq:modulatedsigma}. The modulation factor $\delta'(y')$ is expanded as a Fourier series for each value of $w_0$ considered (the first Fourier component sufficed to accurately describe the grating in the cases considered) and this spatial part multiplies $\sigma(\omega)$ as given by Eqs.~\ref{grapheneConductivity1} and \ref{grapheneConductivity2}. Transmittance and reflectance through the structure are measured in simulations of a single unit cell with periodic boundary conditions.


\begin{acknowledgement}
Emanuele Galiffi was supported through a studentship in the Centre for Doctoral Training on Theory and Simulation of Materials at Imperial College London funded by the EPSRC (EP/L015579/1). John .B. Pendry acknowledges support from the Gordon and Betty Moore foundation. Paloma A. Huidobro acknowledges funding from a Marie Sklodowska-Curie Fellowship within the Horizons 2020 framework.
\end{acknowledgement}

%\begin{acknowledgement}
%
%Please use ``The authors thank \ldots'' rather than ``The
%authors would like to thank \ldots''.
%
%The author thanks Mats Dahlgren for version one of \textsf{achemso},
%and Donald Arseneau for the code taken from \textsf{cite} to move
%citations after punctuation. Many users have provided feedback on the
%class, which is reflected in all of the different demonstrations
%shown in this document.
%
%\end{acknowledgement}

\begin{suppinfo}
Generating a singular grating from an infinite slab \textit{via} a conformal map; discussion on graphene's conductivity close to the singular limit; field plots for a conventional grating; absorption profiles for grating of different periodicities and relative conductivity/chemical potential relation at their working frequencies. 
\end{suppinfo}

% \begin{figure*}[b]
% \includegraphics[width = \textwidth]{Figures/toc_figure.eps}
% \caption{For table of contents/abstract only}
% \end{figure*}


\bibliography{Plasmonics,Graphene_Plasmonics,TO_Metamaterials,books}

\providecommand{\latin}[1]{#1}
\providecommand*\mcitethebibliography{\thebibliography}
\csname @ifundefined\endcsname{endmcitethebibliography}
  {\let\endmcitethebibliography\endthebibliography}{}
\begin{mcitethebibliography}{46}
\providecommand*\natexlab[1]{#1}
\providecommand*\mciteSetBstSublistMode[1]{}
\providecommand*\mciteSetBstMaxWidthForm[2]{}
\providecommand*\mciteBstWouldAddEndPuncttrue
  {\def\EndOfBibitem{\unskip.}}
\providecommand*\mciteBstWouldAddEndPunctfalse
  {\let\EndOfBibitem\relax}
\providecommand*\mciteSetBstMidEndSepPunct[3]{}
\providecommand*\mciteSetBstSublistLabelBeginEnd[3]{}
\providecommand*\EndOfBibitem{}
\mciteSetBstSublistMode{f}
\mciteSetBstMaxWidthForm{subitem}{(\alph{mcitesubitemcount})}
\mciteSetBstSublistLabelBeginEnd
  {\mcitemaxwidthsubitemform\space}
  {\relax}
  {\relax}

\bibitem[Bonaccorso \latin{et~al.}(2010)Bonaccorso, Sun, Hasan, and
  Ferrari]{Bonaccorso2010}
Bonaccorso,~F.; Sun,~Z.; Hasan,~T.; Ferrari,~A.~C. \emph{Nature Photonics}
  \textbf{2010}, \emph{4}, 611--622\relax
\mciteBstWouldAddEndPuncttrue
\mciteSetBstMidEndSepPunct{\mcitedefaultmidpunct}
{\mcitedefaultendpunct}{\mcitedefaultseppunct}\relax
\EndOfBibitem
\bibitem[Vafek(2006)]{Vafek2006}
Vafek,~O. \emph{Physical Review Letters} \textbf{2006}, \emph{97}, 266406\relax
\mciteBstWouldAddEndPuncttrue
\mciteSetBstMidEndSepPunct{\mcitedefaultmidpunct}
{\mcitedefaultendpunct}{\mcitedefaultseppunct}\relax
\EndOfBibitem
\bibitem[Hanson(2008)]{Hanson2008}
Hanson,~G.~W. \emph{Journal of Applied Physics} \textbf{2008}, \emph{103},
  064302\relax
\mciteBstWouldAddEndPuncttrue
\mciteSetBstMidEndSepPunct{\mcitedefaultmidpunct}
{\mcitedefaultendpunct}{\mcitedefaultseppunct}\relax
\EndOfBibitem
\bibitem[Jablan \latin{et~al.}(2009)Jablan, Buljan, and
  Solja{\v{c}}i{\'{c}}]{Jablan2009}
Jablan,~M.; Buljan,~H.; Solja{\v{c}}i{\'{c}},~M. \emph{Physical Review B}
  \textbf{2009}, \emph{80}, 245435\relax
\mciteBstWouldAddEndPuncttrue
\mciteSetBstMidEndSepPunct{\mcitedefaultmidpunct}
{\mcitedefaultendpunct}{\mcitedefaultseppunct}\relax
\EndOfBibitem
\bibitem[Dubinov \latin{et~al.}(2011)Dubinov, Aleshkin, Mitin, Otsuji, and
  Ryzhii]{Dubinov2011}
Dubinov,~A.~A.; Aleshkin,~V.~Y.; Mitin,~V.; Otsuji,~T.; Ryzhii,~V.
  \emph{Journal of Phys.: Condensed Matter} \textbf{2011}, \emph{23},
  145302\relax
\mciteBstWouldAddEndPuncttrue
\mciteSetBstMidEndSepPunct{\mcitedefaultmidpunct}
{\mcitedefaultendpunct}{\mcitedefaultseppunct}\relax
\EndOfBibitem
\bibitem[Koppens \latin{et~al.}(2011)Koppens, Chang, and {Garc{\'{i}}a de
  Abajo}]{Koppens2011}
Koppens,~F. H.~L.; Chang,~D.~E.; {Garc{\'{i}}a de Abajo},~F.~J. \emph{Nano
  letters} \textbf{2011}, \emph{11}, 3370--3377\relax
\mciteBstWouldAddEndPuncttrue
\mciteSetBstMidEndSepPunct{\mcitedefaultmidpunct}
{\mcitedefaultendpunct}{\mcitedefaultseppunct}\relax
\EndOfBibitem
\bibitem[Nikitin \latin{et~al.}(2011)Nikitin, Guinea, Garc{\'{i}}a-Vidal, and
  Mart{\'{i}}n-Moreno]{Nikitin2011}
Nikitin,~A.~Y.; Guinea,~F.; Garc{\'{i}}a-Vidal,~F.~J.; Mart{\'{i}}n-Moreno,~L.
  \emph{Physical Review B} \textbf{2011}, \emph{84}, 195446\relax
\mciteBstWouldAddEndPuncttrue
\mciteSetBstMidEndSepPunct{\mcitedefaultmidpunct}
{\mcitedefaultendpunct}{\mcitedefaultseppunct}\relax
\EndOfBibitem
\bibitem[Fei \latin{et~al.}(2011)Fei, Andreev, Bao, Zhang, {S McLeod}, Wang,
  Stewart, Zhao, Dominguez, Thiemens, Fogler, Tauber, Castro-Neto, Lau,
  Keilmann, and Basov]{Fei2011}
Fei,~Z. \latin{et~al.}  \emph{Nano letters} \textbf{2011}, \emph{11},
  4701--5\relax
\mciteBstWouldAddEndPuncttrue
\mciteSetBstMidEndSepPunct{\mcitedefaultmidpunct}
{\mcitedefaultendpunct}{\mcitedefaultseppunct}\relax
\EndOfBibitem
\bibitem[Basov \latin{et~al.}(2016)Basov, Fogler, and {Garcia de
  Abajo}]{Basov2016}
Basov,~D.~N.; Fogler,~M.~M.; {Garcia de Abajo},~F.~J. \emph{Science}
  \textbf{2016}, \emph{354}, aag1992--aag1992\relax
\mciteBstWouldAddEndPuncttrue
\mciteSetBstMidEndSepPunct{\mcitedefaultmidpunct}
{\mcitedefaultendpunct}{\mcitedefaultseppunct}\relax
\EndOfBibitem
\bibitem[Low \latin{et~al.}(2016)Low, Chaves, Caldwell, Kumar, Fang, Avouris,
  Heinz, Guinea, Martin-Moreno, and Koppens]{Low2017}
Low,~T.; Chaves,~A.; Caldwell,~J.~D.; Kumar,~A.; Fang,~N.~X.; Avouris,~P.;
  Heinz,~T.~F.; Guinea,~F.; Martin-Moreno,~L.; Koppens,~F. \emph{Nature
  Materials} \textbf{2016}, \emph{16}, 182--194\relax
\mciteBstWouldAddEndPuncttrue
\mciteSetBstMidEndSepPunct{\mcitedefaultmidpunct}
{\mcitedefaultendpunct}{\mcitedefaultseppunct}\relax
\EndOfBibitem
\bibitem[Ni \latin{et~al.}(2016)Ni, Wang, Goldflam, Wagner, Fei, McLeod, Liu,
  Keilmann, {\"{O}}zyilmaz, {Castro Neto}, Hone, Fogler, and Basov]{Ni2016}
Ni,~G.~X.; Wang,~L.; Goldflam,~M.~D.; Wagner,~M.; Fei,~Z.; McLeod,~A.~S.;
  Liu,~M.~K.; Keilmann,~F.; {\"{O}}zyilmaz,~B.; {Castro Neto},~A.~H.; Hone,~J.;
  Fogler,~M.~M.; Basov,~D.~N. \emph{Nature Photonics} \textbf{2016}, \emph{10},
  244--247\relax
\mciteBstWouldAddEndPuncttrue
\mciteSetBstMidEndSepPunct{\mcitedefaultmidpunct}
{\mcitedefaultendpunct}{\mcitedefaultseppunct}\relax
\EndOfBibitem
\bibitem[Low and Avouris(2014)Low, and Avouris]{Low2014}
Low,~T.; Avouris,~P. \emph{ACS Nano} \textbf{2014}, \emph{8}, 1086--1101\relax
\mciteBstWouldAddEndPuncttrue
\mciteSetBstMidEndSepPunct{\mcitedefaultmidpunct}
{\mcitedefaultendpunct}{\mcitedefaultseppunct}\relax
\EndOfBibitem
\bibitem[Maier(2007)]{Maier2007Plasmonics}
Maier,~S.~A. \emph{Plasmonics: Fundamentals and Applications}; Springer:
  Boston, MA, 2007\relax
\mciteBstWouldAddEndPuncttrue
\mciteSetBstMidEndSepPunct{\mcitedefaultmidpunct}
{\mcitedefaultendpunct}{\mcitedefaultseppunct}\relax
\EndOfBibitem
\bibitem[Fei \latin{et~al.}(2012)Fei, Rodin, Andreev, Bao, McLeod, Wagner,
  Zhang, Zhao, Thiemens, Dominguez, Fogler, Castro-Neto, Lau, Keilmann, and
  Basov]{Fei2012}
Fei,~Z.; Rodin,~A.~S.; Andreev,~G.~O.; Bao,~W.; McLeod,~A.~S.; Wagner,~M.;
  Zhang,~L.~M.; Zhao,~Z.; Thiemens,~M.; Dominguez,~G.; Fogler,~M.~M.;
  Castro-Neto,~A.~H.; Lau,~C.~N.; Keilmann,~F.; Basov,~D.~N. \emph{Nature}
  \textbf{2012}, \emph{487}, 82--5\relax
\mciteBstWouldAddEndPuncttrue
\mciteSetBstMidEndSepPunct{\mcitedefaultmidpunct}
{\mcitedefaultendpunct}{\mcitedefaultseppunct}\relax
\EndOfBibitem
\bibitem[Chen \latin{et~al.}(2012)Chen, Badioli, Alonso-Gonz{\'{a}}lez,
  Thongrattanasiri, Huth, Osmond, Spasenovi{\'{c}}, Centeno, Pesquera,
  Godignon, Elorza, Camara, {Garc{\'{i}}a de Abajo}, Hillenbrand, and
  Koppens]{Chen2012}
Chen,~J.; Badioli,~M.; Alonso-Gonz{\'{a}}lez,~P.; Thongrattanasiri,~S.;
  Huth,~F.; Osmond,~J.; Spasenovi{\'{c}},~M.; Centeno,~A.; Pesquera,~A.;
  Godignon,~P.; Elorza,~A.~Z.; Camara,~N.; {Garc{\'{i}}a de Abajo},~F.~J.;
  Hillenbrand,~R.; Koppens,~F. H.~L. \emph{Nature} \textbf{2012}, \emph{487},
  77--81\relax
\mciteBstWouldAddEndPuncttrue
\mciteSetBstMidEndSepPunct{\mcitedefaultmidpunct}
{\mcitedefaultendpunct}{\mcitedefaultseppunct}\relax
\EndOfBibitem
\bibitem[Alonso-Gonz{\'{a}}lez \latin{et~al.}(2014)Alonso-Gonz{\'{a}}lez,
  Nikitin, Golmar, Centeno, Pesquera, V{\'{e}}lez, Chen, Navickaite, Koppens,
  Zurutuza, Casanova, Hueso, and Hillenbrand]{Alonso-Gonzalez2014}
Alonso-Gonz{\'{a}}lez,~P.; Nikitin,~a.~Y.; Golmar,~F.; Centeno,~A.;
  Pesquera,~A.; V{\'{e}}lez,~S.; Chen,~J.; Navickaite,~G.; Koppens,~F.;
  Zurutuza,~A.; Casanova,~F.; Hueso,~L.~E.; Hillenbrand,~R. \emph{Science (New
  York, N.Y.)} \textbf{2014}, \emph{344}, 1369--73\relax
\mciteBstWouldAddEndPuncttrue
\mciteSetBstMidEndSepPunct{\mcitedefaultmidpunct}
{\mcitedefaultendpunct}{\mcitedefaultseppunct}\relax
\EndOfBibitem
\bibitem[Alonso-Gonz{\'{a}}lez \latin{et~al.}(2016)Alonso-Gonz{\'{a}}lez,
  Nikitin, Gao, Woessner, Lundeberg, Principi, Forcellini, Yan, V{\'{e}}lez,
  Huber, Watanabe, Taniguchi, Casanova, Hueso, Polini, Hone, Koppens, and
  Hillenbrand]{Alonso-Gonzalez2016}
Alonso-Gonz{\'{a}}lez,~P. \latin{et~al.}  \emph{Nature Nanotechnology}
  \textbf{2016}, \emph{12}, 1--16\relax
\mciteBstWouldAddEndPuncttrue
\mciteSetBstMidEndSepPunct{\mcitedefaultmidpunct}
{\mcitedefaultendpunct}{\mcitedefaultseppunct}\relax
\EndOfBibitem
\bibitem[Zhan \latin{et~al.}(2012)Zhan, Zhao, Hu, Liu, and Zi]{Zhan2012}
Zhan,~T.~R.; Zhao,~F.~Y.; Hu,~X.~H.; Liu,~X.~H.; Zi,~J. \emph{Physical Review
  B} \textbf{2012}, \emph{86}, 165416 (2012)\relax
\mciteBstWouldAddEndPuncttrue
\mciteSetBstMidEndSepPunct{\mcitedefaultmidpunct}
{\mcitedefaultendpunct}{\mcitedefaultseppunct}\relax
\EndOfBibitem
\bibitem[Nikitin \latin{et~al.}(2012)Nikitin, Guinea, Garcia-Vidal, and
  Martin-Moreno]{Nikitin2012b}
Nikitin,~A.~Y.; Guinea,~F.; Garcia-Vidal,~F.~J.; Martin-Moreno,~L.
  \emph{Physical Review B} \textbf{2012}, \emph{85}, 081405(R)\relax
\mciteBstWouldAddEndPuncttrue
\mciteSetBstMidEndSepPunct{\mcitedefaultmidpunct}
{\mcitedefaultendpunct}{\mcitedefaultseppunct}\relax
\EndOfBibitem
\bibitem[Nikitin \latin{et~al.}(2012)Nikitin, Guinea, and
  Martin-Moreno]{Nikitin2012}
Nikitin,~A.~Y.; Guinea,~F.; Martin-Moreno,~L. \emph{Applied Physics Letters}
  \textbf{2012}, \emph{101}, 151119\relax
\mciteBstWouldAddEndPuncttrue
\mciteSetBstMidEndSepPunct{\mcitedefaultmidpunct}
{\mcitedefaultendpunct}{\mcitedefaultseppunct}\relax
\EndOfBibitem
\bibitem[Thongrattanasiri \latin{et~al.}(2012)Thongrattanasiri, Koppens, and
  {Garc{\'{i}}a De Abajo}]{Thongrattanasiri2012a}
Thongrattanasiri,~S.; Koppens,~F. H.~L.; {Garc{\'{i}}a De Abajo},~F.~J.
  \emph{Physical Review Letters} \textbf{2012}, \emph{108}, 047401\relax
\mciteBstWouldAddEndPuncttrue
\mciteSetBstMidEndSepPunct{\mcitedefaultmidpunct}
{\mcitedefaultendpunct}{\mcitedefaultseppunct}\relax
\EndOfBibitem
\bibitem[Slipchenko \latin{et~al.}(2013)Slipchenko, Nesterov, Martin-Moreno,
  and Nikitin]{Slipchenko2013}
Slipchenko,~T.~M.; Nesterov,~M.~L.; Martin-Moreno,~L.; Nikitin,~a.~Y.
  \emph{Journal of Optics} \textbf{2013}, \emph{15}, 114008\relax
\mciteBstWouldAddEndPuncttrue
\mciteSetBstMidEndSepPunct{\mcitedefaultmidpunct}
{\mcitedefaultendpunct}{\mcitedefaultseppunct}\relax
\EndOfBibitem
\bibitem[Peres \latin{et~al.}(2013)Peres, Bludov, Ferreira, and
  Vasilevskiy]{Peres2013}
Peres,~N. M.~R.; Bludov,~Y.~V.; Ferreira,~A.; Vasilevskiy,~M.~I. \emph{Journal
  of Physics: Condensed Matter} \textbf{2013}, \emph{25}, 125303\relax
\mciteBstWouldAddEndPuncttrue
\mciteSetBstMidEndSepPunct{\mcitedefaultmidpunct}
{\mcitedefaultendpunct}{\mcitedefaultseppunct}\relax
\EndOfBibitem
\bibitem[Stauber \latin{et~al.}(2014)Stauber, G{\'{o}}mez-Santos, and {De
  Abajo}]{Stauber2014a}
Stauber,~T.; G{\'{o}}mez-Santos,~G.; {De Abajo},~F.~J. \emph{Physical Review
  Letters} \textbf{2014}, \emph{112}, 077401\relax
\mciteBstWouldAddEndPuncttrue
\mciteSetBstMidEndSepPunct{\mcitedefaultmidpunct}
{\mcitedefaultendpunct}{\mcitedefaultseppunct}\relax
\EndOfBibitem
\bibitem[Miao \latin{et~al.}(2015)Miao, Wu, Li, He, Ding, An, Zhang, and
  Zhou]{Miao2015}
Miao,~Z.; Wu,~Q.; Li,~X.; He,~Q.; Ding,~K.; An,~Z.; Zhang,~Y.; Zhou,~L.
  \emph{Physical Review X} \textbf{2015}, \emph{5}, 041027\relax
\mciteBstWouldAddEndPuncttrue
\mciteSetBstMidEndSepPunct{\mcitedefaultmidpunct}
{\mcitedefaultendpunct}{\mcitedefaultseppunct}\relax
\EndOfBibitem
\bibitem[Zhao and Zhang(2015)Zhao, and Zhang]{Zhao2015}
Zhao,~B.; Zhang,~Z.~M. \emph{ACS Photonics} \textbf{2015}, \emph{2},
  1611--1618\relax
\mciteBstWouldAddEndPuncttrue
\mciteSetBstMidEndSepPunct{\mcitedefaultmidpunct}
{\mcitedefaultendpunct}{\mcitedefaultseppunct}\relax
\EndOfBibitem
\bibitem[Ju \latin{et~al.}(2011)Ju, Geng, Horng, Girit, Martin, Hao, Bechtel,
  Liang, Zettl, Shen, and Wang]{Ju2011}
Ju,~L.; Geng,~B.; Horng,~J.; Girit,~C.; Martin,~M.; Hao,~Z.; Bechtel,~H.~A.;
  Liang,~X.; Zettl,~A.; Shen,~Y.~R.; Wang,~F. \emph{Nature nanotechnology}
  \textbf{2011}, \emph{6}, 630--634\relax
\mciteBstWouldAddEndPuncttrue
\mciteSetBstMidEndSepPunct{\mcitedefaultmidpunct}
{\mcitedefaultendpunct}{\mcitedefaultseppunct}\relax
\EndOfBibitem
\bibitem[Tassin \latin{et~al.}(2013)Tassin, Koschny, and Soukoulis]{Tassin2013}
Tassin,~P.; Koschny,~T.; Soukoulis,~C.~M. \emph{Science} \textbf{2013},
  \emph{341}, 620--621\relax
\mciteBstWouldAddEndPuncttrue
\mciteSetBstMidEndSepPunct{\mcitedefaultmidpunct}
{\mcitedefaultendpunct}{\mcitedefaultseppunct}\relax
\EndOfBibitem
\bibitem[Fan \latin{et~al.}(2015)Fan, Shen, Koschny, and Soukoulis]{Fan2015}
Fan,~Y.; Shen,~N.-H.; Koschny,~T.; Soukoulis,~C.~M. \emph{ACS Photonics}
  \textbf{2015}, \emph{2}, 151--156\relax
\mciteBstWouldAddEndPuncttrue
\mciteSetBstMidEndSepPunct{\mcitedefaultmidpunct}
{\mcitedefaultendpunct}{\mcitedefaultseppunct}\relax
\EndOfBibitem
\bibitem[Li \latin{et~al.}(2015)Li, Kita, Mu{\~{n}}oz, Reshef, Vulis, and
  Yin]{Li2015}
Li,~A.~Y.; Kita,~S.; Mu{\~{n}}oz,~P.; Reshef,~O.; Vulis,~D.~I.; Yin,~M.
  \textbf{2015}, \emph{9}, 738--743\relax
\mciteBstWouldAddEndPuncttrue
\mciteSetBstMidEndSepPunct{\mcitedefaultmidpunct}
{\mcitedefaultendpunct}{\mcitedefaultseppunct}\relax
\EndOfBibitem
\bibitem[Huidobro \latin{et~al.}(2016)Huidobro, Kraft, Maier, and
  Pendry]{Huidobro2016a}
Huidobro,~P.~A.; Kraft,~M.; Maier,~S.~A.; Pendry,~J.~B. \emph{ACS Nano}
  \textbf{2016}, \emph{10}, 5499--5506\relax
\mciteBstWouldAddEndPuncttrue
\mciteSetBstMidEndSepPunct{\mcitedefaultmidpunct}
{\mcitedefaultendpunct}{\mcitedefaultseppunct}\relax
\EndOfBibitem
\bibitem[Pendry \latin{et~al.}(2017)Pendry, Huidobro, Luo, and
  Galiffi]{Pendry2017}
Pendry,~J.~B.; Huidobro,~P.~A.; Luo,~Y.; Galiffi,~E. \emph{Science - accepted}
  \textbf{2017}, \relax
\mciteBstWouldAddEndPunctfalse
\mciteSetBstMidEndSepPunct{\mcitedefaultmidpunct}
{}{\mcitedefaultseppunct}\relax
\EndOfBibitem
\bibitem[Luo \latin{et~al.}(2010)Luo, Pendry, and Aubry]{Luo2010}
Luo,~Y.; Pendry,~J.~B.; Aubry,~A. \emph{Nano letters} \textbf{2010}, \emph{10},
  4186--91\relax
\mciteBstWouldAddEndPuncttrue
\mciteSetBstMidEndSepPunct{\mcitedefaultmidpunct}
{\mcitedefaultendpunct}{\mcitedefaultseppunct}\relax
\EndOfBibitem
\bibitem[Ward and Pendry(1996)Ward, and Pendry]{Ward1996}
Ward,~A.~J.; Pendry,~J.~B. \emph{Journal of Modern Optics} \textbf{1996},
  \emph{43}, 773 -- 793\relax
\mciteBstWouldAddEndPuncttrue
\mciteSetBstMidEndSepPunct{\mcitedefaultmidpunct}
{\mcitedefaultendpunct}{\mcitedefaultseppunct}\relax
\EndOfBibitem
\bibitem[Pendry \latin{et~al.}(2006)Pendry, Schurig, and Smith]{Pendry2006}
Pendry,~J.~B.; Schurig,~D.; Smith,~D.~R. \emph{Science (New York, N.Y.)}
  \textbf{2006}, \emph{312}, 1780--2\relax
\mciteBstWouldAddEndPuncttrue
\mciteSetBstMidEndSepPunct{\mcitedefaultmidpunct}
{\mcitedefaultendpunct}{\mcitedefaultseppunct}\relax
\EndOfBibitem
\bibitem[Leonhardt(2006)]{Leonhardt2006}
Leonhardt,~U. \emph{Science} \textbf{2006}, \emph{312}, 1777\relax
\mciteBstWouldAddEndPuncttrue
\mciteSetBstMidEndSepPunct{\mcitedefaultmidpunct}
{\mcitedefaultendpunct}{\mcitedefaultseppunct}\relax
\EndOfBibitem
\bibitem[Kraft \latin{et~al.}(2015)Kraft, Luo, Maier, and Pendry]{Kraft2015}
Kraft,~M.; Luo,~Y.; Maier,~S.~A.; Pendry,~J.~B. \emph{Physical Review X}
  \textbf{2015}, \emph{5}, 031029\relax
\mciteBstWouldAddEndPuncttrue
\mciteSetBstMidEndSepPunct{\mcitedefaultmidpunct}
{\mcitedefaultendpunct}{\mcitedefaultseppunct}\relax
\EndOfBibitem
\bibitem[Huidobro \latin{et~al.}(2016)Huidobro, Kraft, Kun, Maier, and
  Pendry]{Huidobro2016}
Huidobro,~P.~A.; Kraft,~M.; Kun,~R.; Maier,~S.~A.; Pendry,~J.~B. \emph{Journal
  of Optics} \textbf{2016}, \emph{18}, 44024\relax
\mciteBstWouldAddEndPuncttrue
\mciteSetBstMidEndSepPunct{\mcitedefaultmidpunct}
{\mcitedefaultendpunct}{\mcitedefaultseppunct}\relax
\EndOfBibitem
\bibitem[{Arroyo Huidobro} \latin{et~al.}(2017){Arroyo Huidobro}, Maier, and
  Pendry]{Huidobro2017}
{Arroyo Huidobro},~P.; Maier,~S.~A.; Pendry,~J.~B. \emph{EPJ Applied
  Metamaterials} \textbf{2017}, \emph{4}, 6\relax
\mciteBstWouldAddEndPuncttrue
\mciteSetBstMidEndSepPunct{\mcitedefaultmidpunct}
{\mcitedefaultendpunct}{\mcitedefaultseppunct}\relax
\EndOfBibitem
\bibitem[Savage \latin{et~al.}(2012)Savage, Hawkeye, Esteban, Borisov,
  Aizpurua, and Baumberg]{Savage2012}
Savage,~K.~J.; Hawkeye,~M.~M.; Esteban,~R.; Borisov,~A.~G.; Aizpurua,~J.;
  Baumberg,~J.~J. \emph{Nature} \textbf{2012}, \emph{491}, 574--7\relax
\mciteBstWouldAddEndPuncttrue
\mciteSetBstMidEndSepPunct{\mcitedefaultmidpunct}
{\mcitedefaultendpunct}{\mcitedefaultseppunct}\relax
\EndOfBibitem
\bibitem[Cirac{\`{i}} \latin{et~al.}(2012)Cirac{\`{i}}, Hill, Mock, Urzhumov,
  Fern{\'{a}}ndez-Dom{\'{i}}nguez, Maier, Pendry, Chilkoti, and
  Smith]{Ciraci2012}
Cirac{\`{i}},~C.; Hill,~R.~T.; Mock,~J.~J.; Urzhumov,~Y.;
  Fern{\'{a}}ndez-Dom{\'{i}}nguez,~A.~I.; Maier,~S.~A.; Pendry,~J.~B.;
  Chilkoti,~a.; Smith,~D.~R. \emph{Science (New York, N.Y.)} \textbf{2012},
  \emph{337}, 1072--4\relax
\mciteBstWouldAddEndPuncttrue
\mciteSetBstMidEndSepPunct{\mcitedefaultmidpunct}
{\mcitedefaultendpunct}{\mcitedefaultseppunct}\relax
\EndOfBibitem
\bibitem[Lundeberg \latin{et~al.}(2017)Lundeberg, Gao, Asgari, Tan, {Van
  Duppen}, Autore, Alonso-Gonzalez, Woessner, Watanabe, Taniguchi, Hillenbrand,
  Hone, Polini, and Koppens]{Lundeberg2017}
Lundeberg,~M.; Gao,~Y.; Asgari,~R.; Tan,~C.; {Van Duppen},~B.; Autore,~M.;
  Alonso-Gonzalez,~P.; Woessner,~A.; Watanabe,~K.; Taniguchi,~T.;
  Hillenbrand,~R.; Hone,~J.; Polini,~M.; Koppens,~F. H.~L. \emph{Science}
  \textbf{2017}, \emph{2735}, 1--10\relax
\mciteBstWouldAddEndPuncttrue
\mciteSetBstMidEndSepPunct{\mcitedefaultmidpunct}
{\mcitedefaultendpunct}{\mcitedefaultseppunct}\relax
\EndOfBibitem
\bibitem[Tonouchi(2007)]{Tonouchi2007}
Tonouchi,~M. \emph{Nature Photonics} \textbf{2007}, \emph{1}, 97--105\relax
\mciteBstWouldAddEndPuncttrue
\mciteSetBstMidEndSepPunct{\mcitedefaultmidpunct}
{\mcitedefaultendpunct}{\mcitedefaultseppunct}\relax
\EndOfBibitem
\bibitem[Wunsch \latin{et~al.}(2006)Wunsch, Stauber, Sols, and
  Guinea]{wunsch2006}
Wunsch,~B.; Stauber,~T.; Sols,~F.; Guinea,~F. \emph{New Journal of Physics}
  \textbf{2006}, \emph{8}, 318--318\relax
\mciteBstWouldAddEndPuncttrue
\mciteSetBstMidEndSepPunct{\mcitedefaultmidpunct}
{\mcitedefaultendpunct}{\mcitedefaultseppunct}\relax
\EndOfBibitem
\bibitem[Huidobro \latin{et~al.}(2017)Huidobro, Chang, Kraft, and
  Pendry]{Huidobro2017a}
Huidobro,~P.~A.; Chang,~Y.~H.; Kraft,~M.; Pendry,~J.~B. \emph{Physical Review
  B} \textbf{2017}, \emph{95}, 1--8\relax
\mciteBstWouldAddEndPuncttrue
\mciteSetBstMidEndSepPunct{\mcitedefaultmidpunct}
{\mcitedefaultendpunct}{\mcitedefaultseppunct}\relax
\EndOfBibitem
\end{mcitethebibliography}

\end{document}